\documentclass[aps,pra,superscriptaddress,twocolumn,showpacs]{revtex4-1}

\pdfoutput=1

\usepackage[utf8]{inputenc}
\usepackage[english]{babel}
\usepackage{setspace}  
\usepackage{graphicx}
\usepackage{subfigure}
\usepackage{amssymb}
\usepackage{amsmath}
\usepackage{amscd}
\usepackage{color}
\usepackage{enumerate}

\usepackage{hyperref}
\hypersetup{
        colorlinks=true,
}

\usepackage{amsthm}

\begin{document}

\newcommand{\BIGOP}[1]{\mathop{\mathchoice%
{\raise-0.22em\hbox{\huge $#1$}}%
{\raise-0.05em\hbox{\Large $#1$}}{\hbox{\large $#1$}}{#1}}}
\newcommand{\bigtimes}{\BIGOP{\times}}

\newcommand{\gv}[1]{\ensuremath{\mbox{\boldmath$ #1 $}}} 
\newcommand{\md}[1]{\mathrm{d}#1\,} 
\renewcommand{\d}[2]{\frac{d #1}{d #2}} 
\newcommand{\dd}[2]{\frac{d^2 #1}{d #2^2}} 
\newcommand{\pd}[2]{\frac{\partial #1}{\partial #2}} 
\newcommand{\pdd}[2]{\frac{\partial^2 #1}{\partial #2^2}} 
\newcommand{\pdc}[3]{\left( \frac{\partial #1}{\partial #2} \right)_{#3}} 
\newcommand{\ket}[1]{\left| #1 \right>} 
\newcommand{\bra}[1]{\left< #1 \right|} 
\newcommand{\braket}[2]{\left< #1 \vphantom{#2} \right| \left. #2 \vphantom{#1} \right>} 
\newcommand{\matrixel}[3]{\left< #1 \vphantom{#2#3} \right| #2 \left| #3 \vphantom{#1#2} \right>} 

\newcommand{\nts}[1]{[\emph{\color{red} #1}]}

\title{Minimizing nonadiabaticities in optical-lattice loading}

\author{Michele Dolfi}
\affiliation{Theoretische Physik, ETH Zurich, 8093 Zurich, Switzerland}

\author{Adrian Kantian}
\affiliation{Nordita, KTH Royal Institute of Technology and Stockholm University, Roslagstullsbacken 23, SE-106 91 Stockholm, Sweden}

\author{Bela Bauer}
\affiliation{Station Q, Microsoft Research, Santa Barbara, California 93106-6105, USA}

\author{Matthias Troyer}
\affiliation{Theoretische Physik, ETH Zurich, 8093 Zurich, Switzerland}

\date{\today}

\begin{abstract}
In the quest to reach lower temperatures of ultra-cold gases in optical lattice experiments, nonadiabaticities during lattice loading are one of the limiting factors that prevent the same low temperatures to be reached as in experiments without lattice. Simulating the loading of a bosonic quantum gas into a one-dimensional optical lattice with and without a trap, we find that the redistribution of atomic density inside a global confining potential is by far the dominant source of heating. Based on these results we propose to adjust the trapping potential during loading to minimize changes to the density distribution.  Our simulations confirm that a very simple linear interpolation of the trapping potential during loading already significantly decreases the heating of a quantum gas and we discuss how loading protocols minimizing density redistributions can be designed.
\end{abstract}





\maketitle

\section{Introduction}
Quantum simulations using ultracold atoms confined in a trap allow many interesting phenomena of interacting quantum many body problems to be studied \cite{Bloch:2005gn,esslinger2010,Bloch2012}, but are faced with a continuous quest for lower and lower temperatures that would allow the observation of the myriad of interesting exotic phenomena observed in condensed matter systems.  Progress in cooling was crucial for the realization of Bose-Einstein condensation~\cite{anderson1995,davis1995}, and the observation of the superfluid to Mott insulator transition both for bosons~\cite{greiner2002} and fermions~\cite{jordens2008,schneider2008}, to name just a few examples. Despite recently observed short-range magnetic correlations \cite{greif2013,hart2015}, the transition to a N\'eel state with long range order has not yet been observed. Other, more exotic, phases of interacting strongly correlated fermions, such as high-temperature superconducivity \cite{orenstein2000}, occur at even lower temperatures. Both in cuprate superconductors and in the Hubbard model \cite{Staar2013}, the superconducting transition temperature is more than ten times lower than the scale of antiferromagnetic ordering.  

While cooling of fermionic quantum gases has reached temperatures as low as $T/E_F\approx 0.05$ in the continuum \cite{zwierlein2006}, it has been harder to achieve low temperatures in lattice experiments. In particular, ramping up the optical lattice potential cannot practically happen fully adiabatically and the temperature of the gas increases substantially during loading~\cite{esslinger2010}. Ramping up the lattice more slowly to get closer to adiabaticity is not expedient either, since the quantum gas heats up over time due to spontaneous emission from the optical lattice \cite{trotzky2010,pichler2010}. Adding compensating beams allows evaporating cooling also in the lattice~\cite{mathy2012}, but experimentally achieving lower temperatures is still an open challenge.

\begin{figure}[b]
\centering
\includegraphics[width=.95\linewidth]{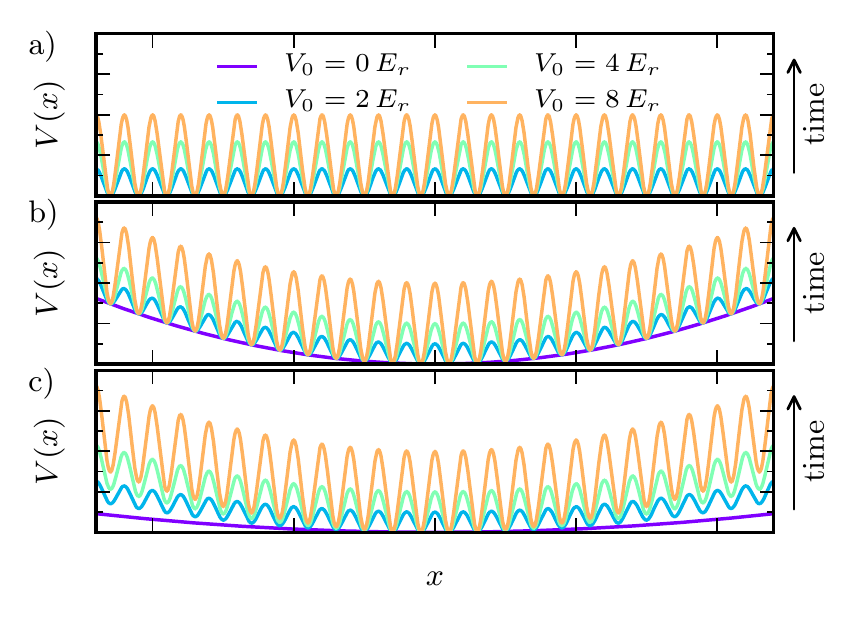}
\caption{(Color online) Sketch of optical lattice loading. We show the potential at various times corresponding to lattice strengths $V_0$ ranging from 0 to $8E_R$, illustrating how the lattice is ramped up.   a) in a homogeneous system; b) in a constant trapping potential and c) while changing the trap to minimize density redistributions.}
\label{fig:ramp schematic}
\end{figure}

Given this challenges, finding a way to reduce heating due to unavoidable non-adiabatic lattice loading would be highly welcome  to reach lower temperatures than are accessible today. In this paper we explore strategies to achieve this goal by numerically simulating the loading of a one-dimensional Bose gas into an optical lattice. Our main result  is that neither non-adiabatic loading into higher bands of the optical lattice, nor the crossing of the phase transition to a Mott insulator are the dominant sources of heating. These effects are more than an order of magnitude smaller than the heating due to redistributing the atoms in the trap. Adjusting the trapping potential during loading, which is easy to achieve experimentally, can significantly reduce heating during optical lattice loading.

An accurate description of lattice loading needs to start from a continuum model since initially the lattice is turned off. Approximating  the optical lattice by a one-band model~\cite{jaksch1998}, as has been done in Refs.~\cite{zakrzewski2009,bernier2011,bernier2012} is valid only in deep lattices during the last phase of loading. Mean-field approaches for continuum models \cite{haque2013,masuda2014} may not reliably catch excitations above the ground state nor accurately describe the crossing of phase transitions.

In this paper we thus perform numerical simulations for a continuum model that allows us to reliably treat both the shallow lattice regime as well as the strong correlation effects in deep optical lattices in a controlled fashion. Specifically, we study $N$ interacting bosons in one dimension described by the Hamiltonian
\begin{multline}
\label{eq:hamiltonian}
\mathcal{H} = \int_0^{L}\md{x} \hat\psi^\dagger(x) \left[ -\frac{\hbar^2}{2m}\dd{}{x} + V(x) \right] \hat\psi(x) \\
 + \frac{g}{2}\int_0^{L}\md{x} \hat\psi^\dagger(x) \hat\psi^\dagger(x) \hat\psi(x) \hat\psi(x),
\end{multline}
where the external potential due to the optical lattice and an external confinement with frequency $\omega$ is given by
\begin{equation}
\label{eq:potential}
V(x) = V_0 \cos^2 ( 2\pi / \lambda \cdot x) + \frac{1}{2} m \omega^2 x^2.
\end{equation}
The field operator $\hat\psi^\dagger(x)$ creates a boson with mass $m$ (we will consider $^{87}$Rb atoms) at position $x$ , $L$ denotes the number of unit cells of size $a=\lambda/2$, where for the laser wavelength $\lambda=826$nm we use the values of the experiment in Ref.~\cite{kohl2004}. The interaction strength $g=2\hbar \omega_\perp a_s$ is determined  by the scattering length $a_s$ and the transverse confining frequency $\omega_\perp$ \cite{olshanii1998}.  Unless noted otherwise we specify energies in terms of the recoil energy $E_r=h^2 /(2m \lambda^2 )$. 

We use a finite difference discretization to simulate the continuum Hamiltonian \eqref{eq:hamiltonian} on a grid with $M$ grid points per unit cell, corresponding to a lattice spacing $\Delta x = a / M$. The continuum model then maps to a lattice Bose-Hubbard model with nearest neighbour hopping $J(\Delta x)=(\hbar^2 / 2m) / \Delta x^2$, on-site interaction $U(\Delta x)=g / \Delta x$ and a site-dependent chemical potential $V_i(\Delta x) = V(\Delta x/2 + i\Delta x) + 2 (\hbar^2 / 2m) / \Delta x^2$. Note that this model is different from the effective Hubbard model for deep lattices \cite{jaksch1998}, since it consists of $M$ lattice sites per unit cell and explicitly includes the optical lattice potential.

\begin{figure}[t]
\centering
\includegraphics[width=\columnwidth]{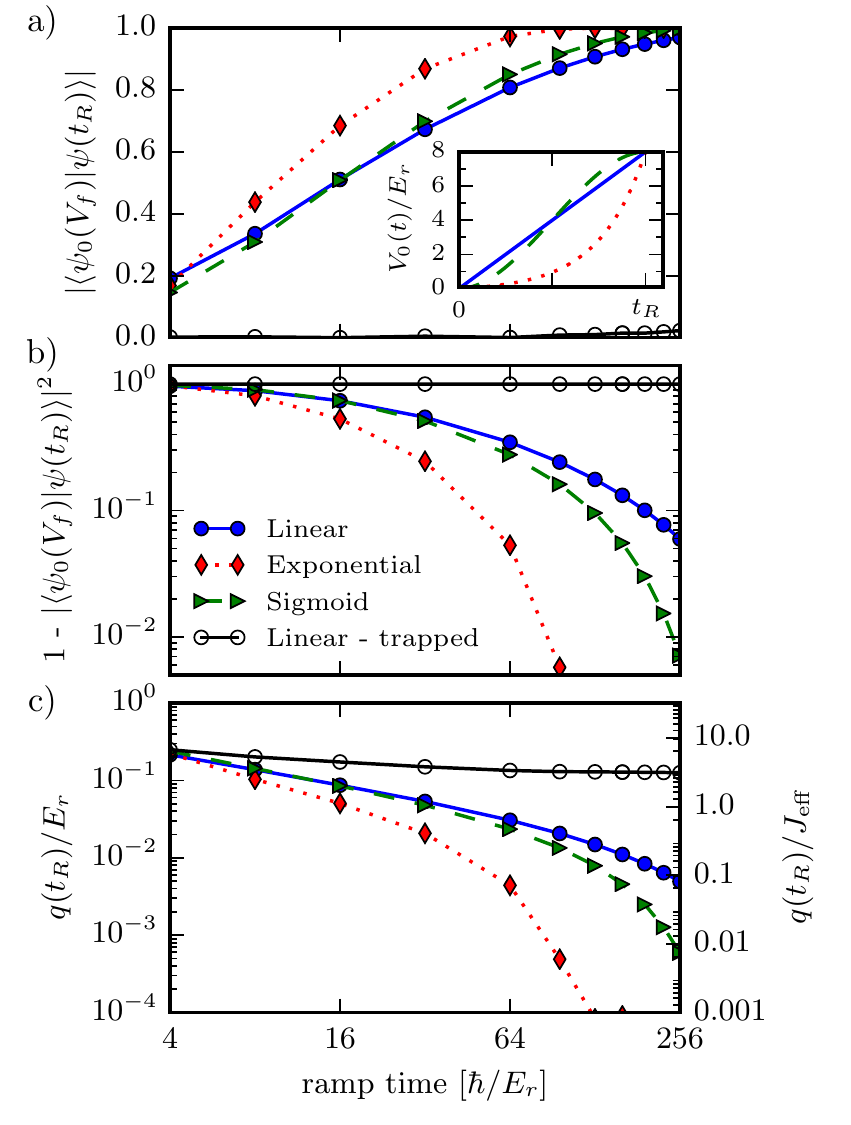}
\caption{(Color online) Dependence on the ramp time $t_R$ of a) the fidelity of the final state after ramping, b) the same data plotted as the probability of being in an excited state $1 - |\langle \psi_0(V_f) | \psi(t_R) \rangle|^2$ in log-scale - and c) the excess energy per particle $q$ in units of the recoil energy $E_r$ and of the effective final hopping amplitude $J_{\rm eff}$.  The optical lattice is ramped up to a final strength of $V_f=8\, E_r$ at fixed interactions $g=2\, E_r\, \lambda/2$. Results are shown for three different ramp profiles, displayed in the inset of a), and described in Eqns.~\eqref{eq:prof.lin}--\eqref{eq:prof.sramp}. Full symbols refer to homogeneous system with $N=24$ particles in $L=24$ sites with hard wall boundary conditions. Open symbols  refer to an inhomogenous system of the same size but with with $N=12$ particles in a confining harmonic potential with $\omega=0.3\, (\hbar/E_r)^{-1}$.}
\label{fig:results no trap}
\end{figure}

\section{Simulation method}
To solve this model we use the density matrix renormalization group method (DMRG)~\cite{white1992,schollwock2011}, which in one dimension provide excellent approximation for low-energy states, and has also proven to yield very good comparisons to experimental data~\cite{cheneau2012,fukuhara2013}. The standard DMRG approach has serious convergence problems for large dilute lattices that arise from a discretization as described above with small $\Delta x$. To overcome these problems we use the multigrid DMRG algorithm~\cite{dolfi2012} which leads to fast convergence even for very dilute systems on large lattices. In essence, the multigrid DMRG method avoids the convergence issues of dilute systems by first solving the model for a large discretization $\Delta x$. This solution is then used to recursively initialize simulations at  decreasing values of $\Delta x$, down to $\Delta x= a/8$. We use this algorithm both to prepare the initial state of the system in the absence of an optical lattice and to calculate reference ground state wave functions $\ket{\psi_0(V_0)}$ and corresponding energies $E_0(V_0)$ at various optical lattice depths $V_0$. In all simulations we used a discretization of up to $M=8$ grid points per optical lattice site, and found that a bond dimension $D=400$ is large enough to see convergence to the ground state.

Time evolution during optical lattice loading is simulated using  time-dependent variants of DMRG \cite{vidal2003,daley2004,white2004}, employing a second-order Trotter decomposition of the time evolution operator at constant time step $\Delta t=0.01\, \hbar / E_r$ and $D=400$. Starting from the ground state without any optical lattice ($V_0 = 0$), we can simulate the evolution of the wave function $\ket{\psi(t)}$ and its mean energy $E(t)=\matrixel{\psi(t)}{\mathcal{H}(t)}{\psi(t)}$ under arbitrary ramping profiles $V(t)$ up to a final strength $V_f \equiv V_0(t_R)$ at the end of loading at time $t_R$. We calculate, in particular, the excess energy  per particle 
\begin{equation}
q(t) = [E(t) - E_0(V(t))] / N
\end{equation}
 and the fidelity  $|\braket{\psi_0(V_0(t))}{\psi(t)}|$ with respect to the ground state $\ket{\psi_0(V_0)}$ of the instantaneous Hamiltonian.

\section{Homogeneous system}
We start by investigating the effects of nonadiabaticities during optical lattice loading in the absence of a harmonic confinement $\omega$. A one-dimensional optical lattice with bosons at integer fillings undergoes a phase transition from a superfluid phase to a Mott insulator when the strength of the optical lattice $V_0$ is increased~\cite{buchler2003,haller2010}. For our simulations we choose unit filling with $N=L$ and an interaction strength $g=2\, E_r\, \lambda/2$, for which the final state with optical lattice strength $V_f = 8E_r$ is in the Mott insulating phase, with a dimensionless Lieb-Liniger parameter $\gamma = Lmg/ \hbar^2 N  \approx 10$. In the final deep optical lattice the system is well described by an effective  single-band Bose-Hubbard model with a nearest-neighbors hopping amplitude $J_{\rm eff} \approx 0.03\, E_r$ and on-site interaction $U_{\rm eff}/J_{\rm eff} \approx 125$.

\begin{figure}[ht]
\centering
\includegraphics[width=\linewidth]{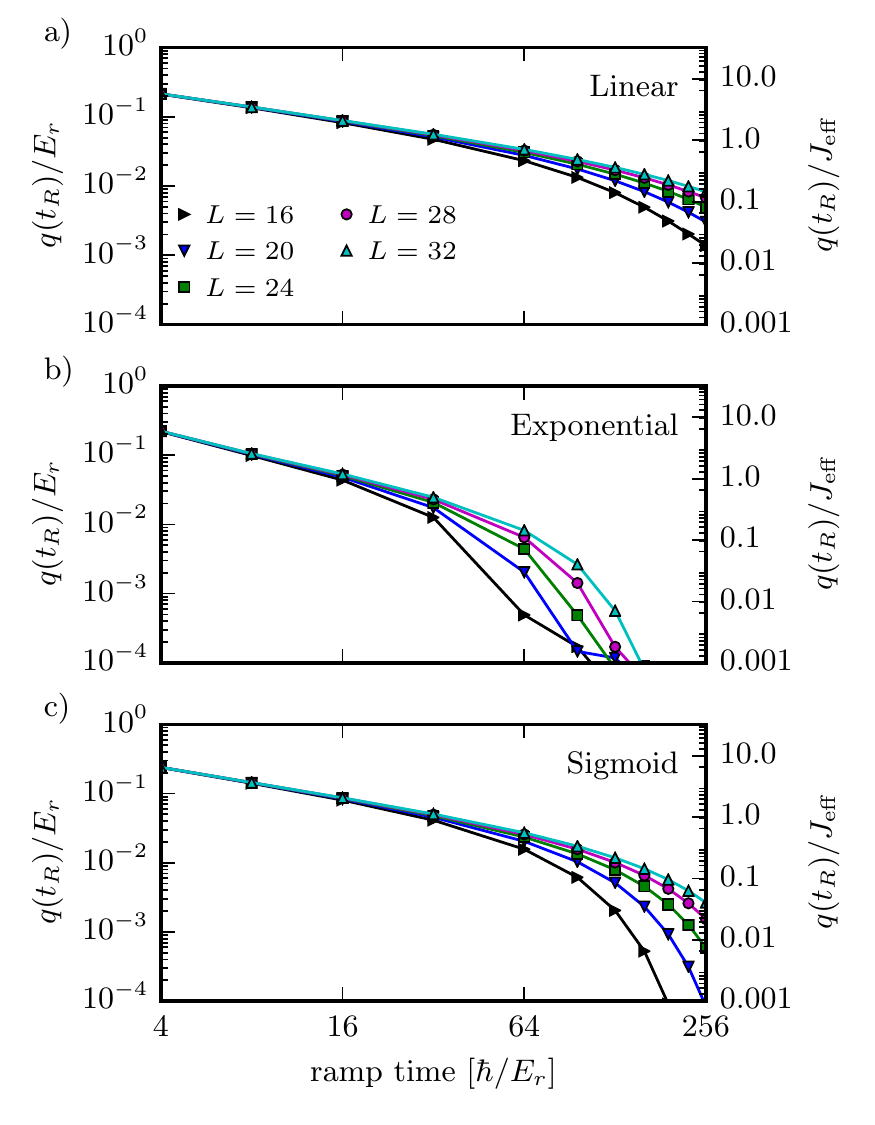}
\caption{(Color online) Dependence of the final excess energy per particle $q(t_R)$ in units of the recoil energy $E_r$ and of the effective final hopping amplitude $J_{\rm eff}$. The optical lattice is ramped up to a final strength of $V_f=8\, E_r$ at fixed interactions $g=2\, E_r\, \lambda/2$ for various system sizes $L=16, 20, 24, 28, 32$ at unit filling. Results in the subpanels show three different ramp profiles a) linear, b) exponential and c) sigmoid, as plotted in Fig.~\ref{fig:results no trap} and described in Eqns.~\eqref{eq:prof.lin}--\eqref{eq:prof.sramp}.}
\label{fig:finite_size}
\end{figure}

Our first goal is to investigate which ramp profile gives minimal heating. We consider in particular the following three ramp shapes:
\begin{eqnarray}
\label{eq:prof.lin}  V_0^{\rm linear}(t) / V_f &=& t / t_R\\
\label{eq:prof.exp}  V_0^{\rm exponential}(t) / V_f &=& \left[ e^{t / \tau} -1\right] / \left[ e^{t_R/\tau} -1\right]\\
\label{eq:prof.sramp}  V_0^{\rm sigmoid}(t) / V_f&=& (t/t_R)^2 \left[-2(t/t_R) + 3\right]
\end{eqnarray}
with $\tau = 0.25\,t_R$. As our results in Fig.~\ref{fig:results no trap} show, the exponential profile, starting with a slow initial turn-on of the lattice leads to the lowest excess energy and highest fidelity. This result can be qualitatively understood by considering the small band gap between bands in weak optical lattices, which requires that care must be taken not to populate higher bands.

Comparing to analytical predictions for the number of defects based on an effective sine-Gordon model \cite{de-grandi2008} we find that the decay of the excess energy $q(t_r)$ is inconsistent with the predicted exponents. Similar discrepancies were previously seen  in numerical simulations of a Bose-Hubbard model \cite{bernier2011}. This indicates that non-universal physics beyond the sine-Gordon model is relevant at experimental ramp speeds and that a numerical simulation of the full model is important.

DMRG methods are performed in finite size systems with open boundary conditions. This might produce finite size effects in our results. Fig.~\ref{fig:finite_size} shows that for all ramp up profiles the finite system size causes the heating to drop significantly for long ramp times $t_R$. With increasing system size the power-law decay is seen for a larger range of ramp times.

\begin{figure*}[t]
\centering
\includegraphics{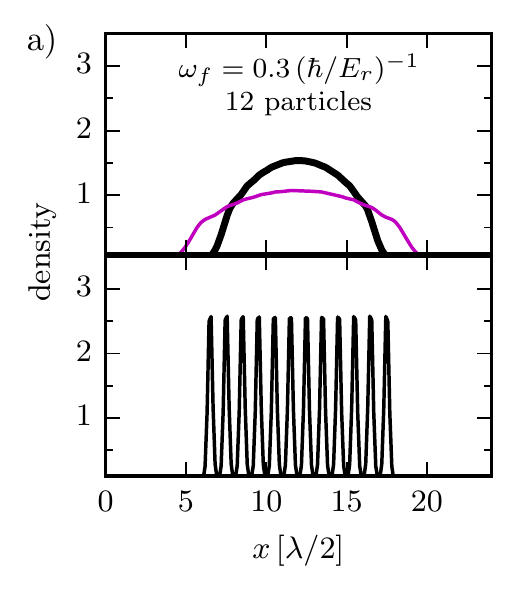}
\includegraphics{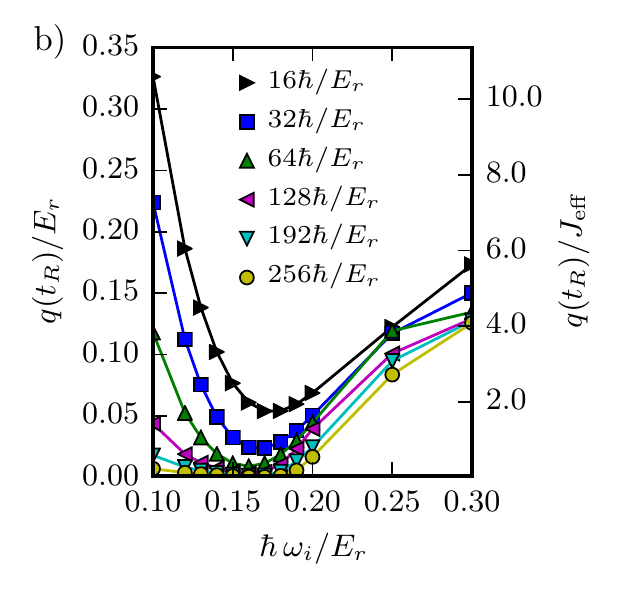}
\includegraphics{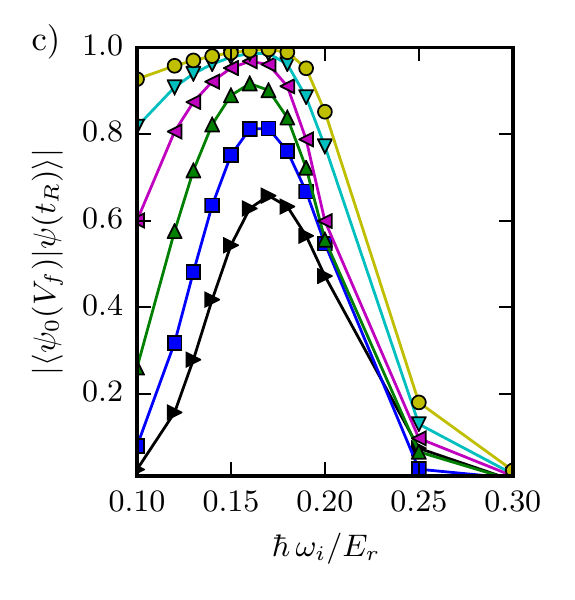}
\caption{(Color online)  a) Local density profile for the initial state with the same trapping frequency $\omega_i=\omega_f=0.3\, (\hbar/E_r)^{-1}$ as the final target state (bold solid line in the upper panel), and for optimal initial state with $\omega_i\approx 0.16\, (\hbar/E_r)^{-1}$ (thin solid line in the upper panel). The final state, a unit filling Mott insulator, is shown in the lower panel. b) Excess energy per particle $q = (E(t_R) - E_0) / N$ at the end of the ramp up as a function of the initial trap frequency $\omega_i$ for various ramp times. c) Fidelity of the final wave function $\ket{\psi(t_R)}$ against the true ground state $\ket{\psi_0(V_f)}=\ket{\psi_0(V_0=V_0(t_R))}$ as a function of the initial trap frequency $\omega_i$ for various ramp times. All calculations are performed on a system of $L=24$ optical lattice sites with $N=12$ particles and targeting a final state with trapping frequency $\omega_f=0.3\, (\hbar/E_r)^{-1}$.}
\label{fig:results_trap_unimott}
\end{figure*}

\section{Trap effects}
Using realistic experimental parameters we find that even with very short loading of only $t_R = 64\, \hbar / E_r \approx 3\,ms$ we only have minimal heating of less than 1\% of $J_{\rm eff}$, far less than observed in experiments. 
While further optimization of the ramp profile will certainly decrease heating further, we do not follow this route since our results already indicate that ramping up the lattice in a homogeneous system cannot be the main source of heating in experiments. 

Repeating the simulation with an added harmonic trapping potential $\omega=0.3\, (\hbar/E_r)^{-1}$, also shown in Fig.~\ref{fig:results no trap}, immediately leads to significantly stronger heating that decreases much more slowly upon increasing $t_R$. This demonstrates that trap effects are the main source of heating during optical lattice loading.

\begin{figure*}[t]
\centering
\includegraphics{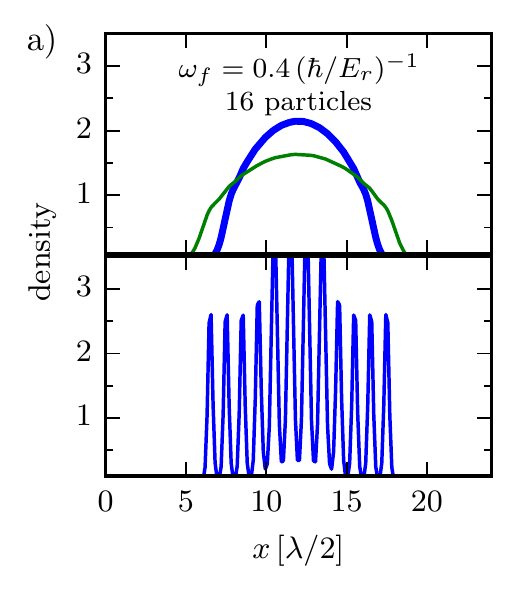}
\includegraphics{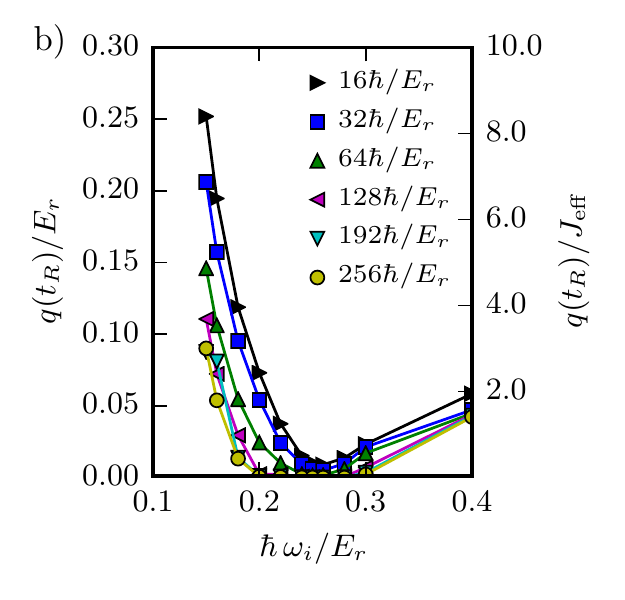}
\includegraphics{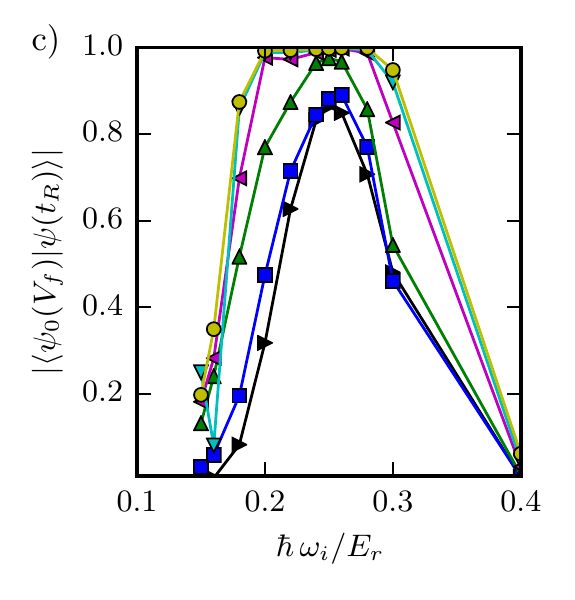}
\caption{(Color online) a) Local density profile for the initial state with the same trapping frequency $\omega_i=\omega_f=0.4\, (\hbar/E_r)^{-1}$ as the final target state (bold solid line in the upper panel), and for optimal initial state with $\omega_i\approx 0.25\, (\hbar/E_r)^{-1}$ (thin solid line in the upper panel). The final state, a Mott insulator with a superfluid core, is shown in the lower panel. b) Excess energy per particle $q = (E(t_R) - E_0) / N$ at the end of the ramp up as a function of the initial trap frequency $\omega_i$ for various ramp times. c) Fidelity of the final wave function $\ket{\psi(t_R)}$ against the true ground state $\ket{\psi_0(V_f)}=\ket{\psi_0(V_0=V_0(t_R))}$ as a function of the initial trap frequency $\omega_i$ for various ramp times. All calculations are performed on a system of $L=24$ optical lattice sites with $N=16$ particles and targeting a final state with trapping frequency $\omega_f=0.4\, (\hbar/E_r)^{-1}$.}
\label{fig:results_trap_doublemott}
\end{figure*}

The main effect of the trapping potential $\omega > 0$ is to modify the homogeneous density distribution to an inhomogeneous one, initially a Gaussian density profile as shown in Figs.~\ref{fig:results_trap_unimott} and~\ref{fig:results_trap_doublemott}. During loading the density distribution changes significantly. We focus our simulations on a linear ramp and two commonly targeted final states: a unit filling Mott insulating core region in Fig.~\ref{fig:results_trap_unimott} obtained with a trapping frequency $\omega=0.3\, (\hbar/E_r)^{-1}$ and a superfluid core with density  larger than one in Fig.~\ref{fig:results_trap_doublemott} obtained with a trapping frequency $\omega=0.4\, (\hbar/E_r)^{-1}$ and more particles.

The redistribution of the atoms from the center towards the edges is the dominant source of non-adiabatic heating during loading. We find that the excess energy $q(t_R)>J_{\rm eff}$ and the fidelity is close to zero even for the longest ramp times (see the right-most data point $\omega_i=\omega_f$ in Figs.~\ref{fig:results_trap_unimott} and~\ref{fig:results_trap_doublemott}).  Significant heating due to density redistribution will occur even for the longest  ramp times  used in experiments.

\begin{figure}[b]
\centering
\includegraphics[width=.95\linewidth]{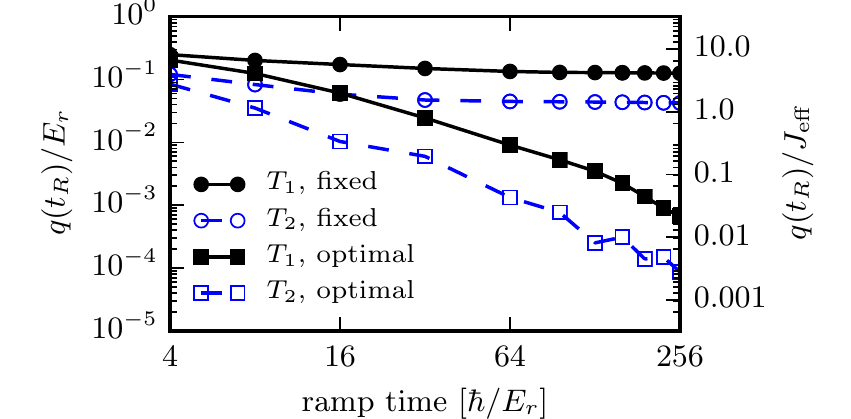}
\caption{(Color online) Dependence on the ramp time $t_R$ of the excess energy per particle. Black solid lines refer to the target state of Fig.~\ref{fig:results_trap_unimott} with fixed trap ($T_1$, fixed) and optimal initial trap ($T_1$, optimal). Blue dashed lines refer to the target state of Fig.~\ref{fig:results_trap_doublemott} with fixed trap ($T_2$, fixed) and optimal initial trap ($T_2$, optimal).}
\label{fig:trap_heating_vs_tR}
\end{figure}

Note that when the physical model contains the trapping harmonic potential, the system adapts its size accordingly. Hence, boundary effects are expected in the actual physical system. Technically, the DMRG simulation is anyway performed on a finite system with open boundary conditions, but the system size is chosen to be larger than the actual physical size, such that the open boundary effects are negligible.

\begin{figure}[b]
\centering
\includegraphics[width=\linewidth]{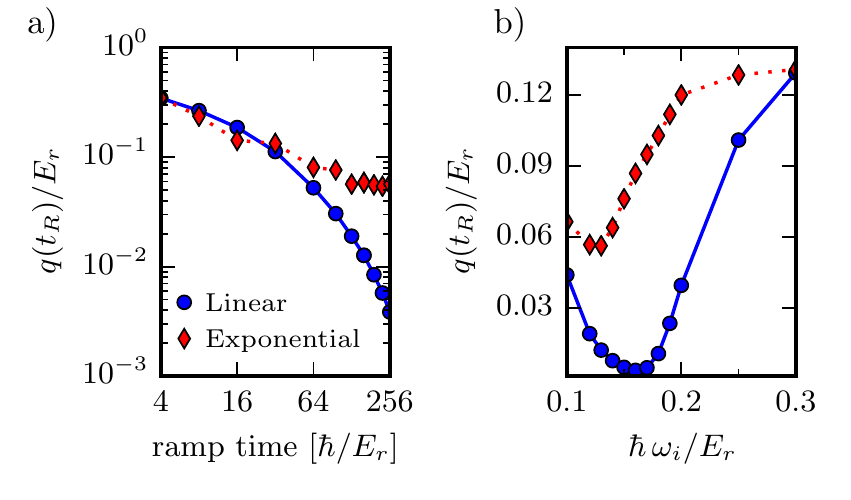}
\caption{(Color online) Comparison of linear (solid blue line) and exponential (dotted red line) lattice loading in a trapped system of length $L=24$ and $N=12$ particles. a) Dependence on the ramp time $t_R$ of the excess energy per particle $q = (E(t_R) - E_0) / N$ in units of the recoil energy $E_r$ for an initial trap frequency $\omega_i = 0.12\, (\hbar/E_r)^{-1}$ which provides the best results for the exponential loading (see panel b) ). b) Excess energy per particle $q$ at the end of the ramp up as a function of the initial trap frequency $\omega_i$ for a ramp time $t_R = 128\, \hbar / E_r$.}
\label{fig:trap_expramp}
\end{figure}

\section{Reducing heating by trap shaping}
The insight that density redistribution of the main source of heating opens a way to significantly reduce heating. We propose  to dynamically adjust the trapping potential during loading to minimize the change in particle distribution. We find that already a very simple protocol, of linearly interpolating the trapping frequency during loading 
\begin{equation}
\omega(t) = \omega_i + \frac{t}{t_R}(\omega_f-\omega_i)
\end{equation}
drastically reduces heating and can easily be implemented experimentally. Calculating heating and fidelity for various values of the initial trapping frequency $\omega_i$ we find a reduction in heating by more than an order of magnitude. We also find that, indeed, heating is minimized if the initial density distribution closely mimics the final one. We show these optimal initial distributions in the left panels of Figs.~\ref{fig:results_trap_unimott} and~\ref{fig:results_trap_doublemott}. Fig.~\ref{fig:trap_heating_vs_tR} shows that the power law decrease of the fidelity and excess energy with $t_R$ that we had seen for a homogeneous system is recovered in the trapped case for an optimal value of the initial frequency of $\omega_i\approx 0.16\, (\hbar/E_r)^{-1}$ and $\omega_i \approx 0.25\, (\hbar/E_r)^{-1}$ for loading into the superfluid and Mott insulating phases, respectively.

The proposed protocol achieves very similar results also for more complex lattice loading profiles as shown in Fig.~\ref{fig:trap_expramp} where we compare the linear and exponential loading combined with the linear change in the trap frequency. In this specific results the linear ramp slightly outperformed the chosen exponential ramp. This is not totally unexpected if one considers that, because of the local varying chemical potential in the trap, there is always a spacial region which is critical, therefore we have to move slow throughout the whole ramp and not only in the beginning.

\section{Discussion and Outlook} 
The strong effects of the density redistribution in a trap compared to loading in a homogeneous lattice can be understood as follows. In the homogeneous system the mean density remains the same and the main effect of ramping up the lattice is to locally change the Wannier functions and the density fluctuations. Since there is a gap between the energy levels within a lattice site and a large overlap matrix element for the evolution within the lowest state this is easy to follow adiabatically. Redistribution in the lattice, however, requires atoms to tunnel across several lattice sites. The matrix elements for tunneling are much smaller and additionally changing the density distribution can easily induce density oscillations. Since these have a much smaller gap (and are gapless in the thermodynamic limit) one thus has to load much slower in the presence of a trap.

While our numerical results were obtained for a one-dimensional system, the finding that density redistribution is the dominant source of heating applies more generally also to higher dimensional optical lattices, fermionic quantum gases and mixtures. While dynamical simulations beyond one dimension are out of reach of current simulation methods, we propose a procedure to find optimized loading protocols based on purely static simulations. Using quantum Monte Carlo (QMC) simulations for bosons in continuum descriptions of weak optical lattices \cite{nguyen2014} or realistically sized lattice models  \cite{trotzky2010}) one can calculate the density profiles to find a sequence $\omega(t)$ of trapping frequencies that minimizes density redistributions. For fermionic systems, QMC results for homogeneous lattice models  \cite{Fuchs:2011ch,Imriska:2014dg,paiva2011}, QMC results for continuum models \cite{pilati2014}, or density functional theory results \cite{Ma:2012kb} can be combined with a local density approximation to similarly obtain density profiles of fermions in a trap and design improved loading strategies.  

As we observed in one-dimensional Bose gases, we expect that also there linearly interpolating the trap from $\omega_i$ to $\omega_f$ chosen such that density redistribution is minimized will significantly decrease heating. Better protocols may be designed by using an optimized form of $\omega(t)$, or by designing anharmonic traps that are able to further reduce density redistributions.  By reducing heating during optical lattice loading interesting phases, such as a N\'eel state with true long range order may already be feasible with current experimental setups.

\begin{acknowledgments}
We thank A. J. Daley, D. Greif, J. Gukelberger, M. Iazzi and L. Wang for enlightening discussions.
The simulations were performed using the ALPS MPS code~\cite{dolfi2014,bauer2011-alps,albuquerque2007} on the  M\"onch cluster of ETH Zurich.
This project was supported by the Swiss National Science Foundation through the National Center of Competence in Research Quantum Science and Technology QSIT and by ERC Advanced Grant SIMCOFE. MT acknowledge hospitality of the Aspen Center for Physics, supported by NSF grant PHY-1066293.
\end{acknowledgments}

\bibliography{bibliography}{}
\bibliographystyle{apsrev4-1}

\end{document}